\documentstyle[12pt,twoside,fleqn,espcrc1,epsfig]{article}

\newcommand{\AmS}{{\protect\the\textfont2
  A\kern-.1667em\lower.5ex\hbox{M}\kern-.125emS}}

\title{Primordial hadrosynthesis in the Little Bang}

\author{Ulrich Heinz\address{Theoretical Physics Division, CERN, 
                             CH-1211 Geneva 23}\thanks{On leave from
        Institut f\"ur Theoretische Physik, Regensburg;
        work supported by BMBF, DFG, and GSI.}
        }

\begin{document}
\maketitle

\begin{flushright}
\vspace*{-6.2cm}
{\footnotesize{\sffamily CERN-TH/99-209\\nucl-th/9907060}}
\end{flushright}
\vspace*{4.7cm}

\begin{abstract}
The present status of soft hadron production in high energy heavy-ion 
collisions is summarized. In spite of strong evidence for extensive 
dynamical evolution and collective expansion of the fireball before 
freeze-out I argue that its chemical composition is hardly changed 
by hadronic final state interactions. The measured hadron yields thus 
reflect the primordial conditions at hadronization. The observed 
production pattern is consistent with statistical hadronization at 
the Hagedorn temperature from a state of uncorrelated, color deconfined 
quarks and antiquarks, but requires non-trivial chemical evolution of 
the fireball in a prehadronic (presumably QGP) stage {\em before} 
hadron formation.
\end{abstract}

\section{Heavy-ion data and the nuclear phase diagram}

\setcounter{footnote}{0}

Relativistic heavy-ion collisions are studied with the goal of
creating hot and dense hadronic matter and to investigate the
nuclear phase diagram at high temperatures and densities, including
the expected phase transition to a color deconfined quark-gluon plasma.
But even if the energy deposited in the reaction zone is quickly 
randomized and the fireball constituents reach an approximate state 
of local thermal equilibrium, a simple connection between heavy-ion 
observables and the phase diagram is still not easy: the pressure 
generated by the thermalization process blows the fireball apart, 
causing a strong time dependence of its thermodynamic 
conditions which is difficult to unfold from the experimental 
observations. There are therefore two fundamental issues to be solved
before one can extract information on the nuclear phase diagram from 
heavy-ion experiments: (1) To what degree does the fireball approach 
local thermal equilibrium? (2) Which observables are sensitive to
which stage(s) of its dynamical evolution, and which is the most 
reliable procedure for extracting the corresponding thermodynamic 
information?

Combining microscopic models for the dynamical fireball evolution 
with macroscopic thermal models for the analysis of heavy-ion data,
significant progress has been recently made in answering both of 
these questions. Crucial for this achievement was the dramatically 
improved quantity and quality of hadron production data from the 
analysis of collisions between very heavy nuclei (Au+Au, Pb+Pb) 
from SIS to SPS. 

Fig.~1 shows a compilation by Cleymans and Redlich \cite{CR99} 
of hadronic freeze-out points in the nuclear phase diagram from 
various collision systems and beam energies. The upper set of 
points, parametrized by a constant average energy per particle 
$\langle E\rangle/\langle N \rangle$=1 GeV \cite{CR99}, 
is obtained from measured hadron yields. They indicate the 
average thermodynamic conditions at {\em chemical freeze-out} when 
the hadron abundances stopped evolving. The lower set of points, 
compared with lines of constant energy and particle density 
\cite{CR99}, is obtained from analyses of hadron momentum spectra 
and/or two-particle momentum correlations. They indicate 
{\em thermal freeze-out}, i.e. the decoupling of the momentum 
distributions. The chemical and thermal freeze-out points at the 
SPS and AGS, respectively, are connected by isentropic expansion 
trajectories with $S/B \approx$~36-38 for the SPS and 12-14 for 
the AGS.

\begin{minipage}[c]{10cm}
\label{F1}
\vspace*{-2.5cm}
\hspace*{-1.5cm}
  \epsfxsize 10cm
  \epsfbox{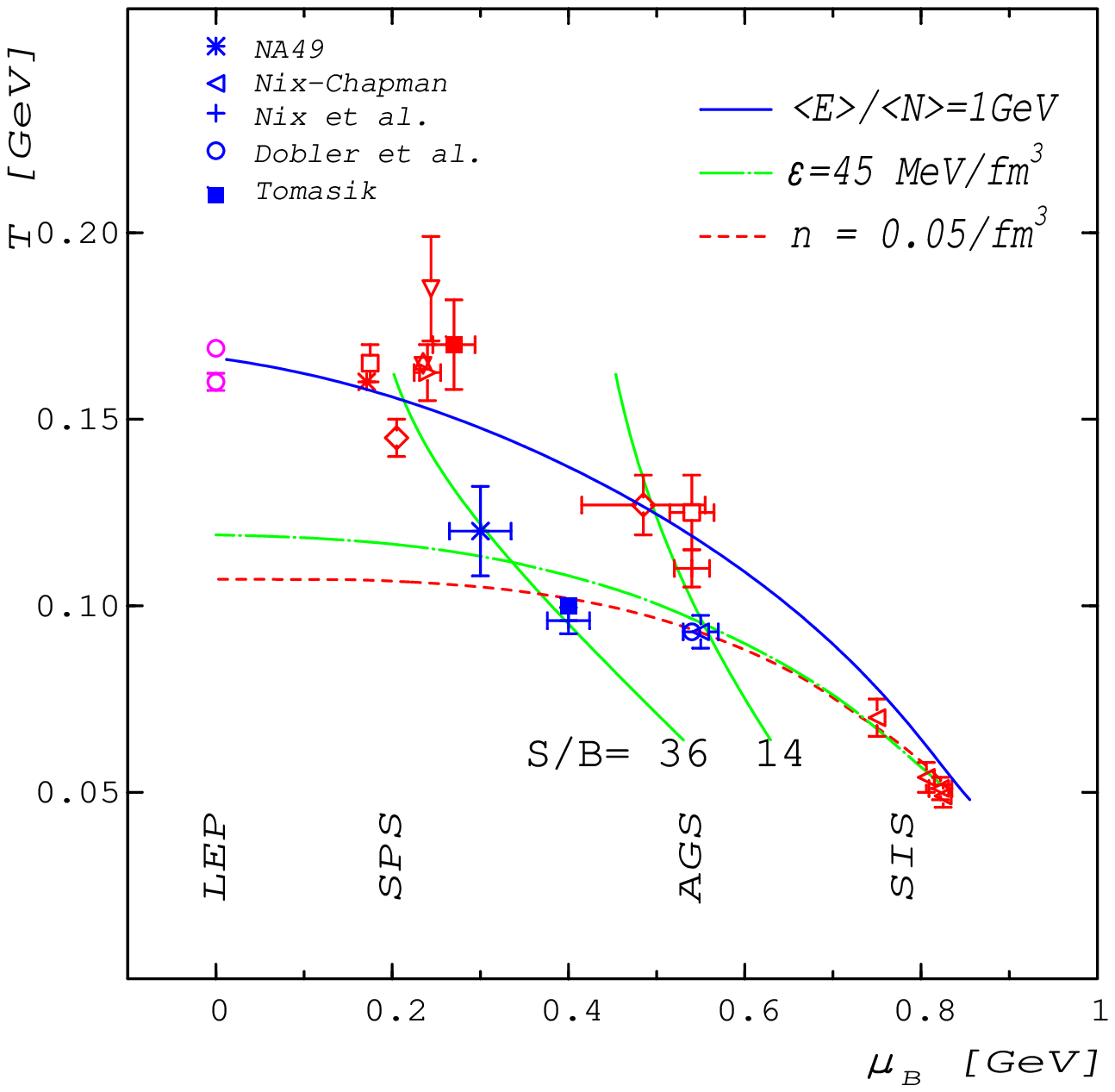}
\end{minipage}
\hfill\hspace*{-3cm}
\parbox[c]{7cm}{\vspace*{-0.5cm}
    Figure~1. Compilation by Cleymans and Redlich \cite{CR99}
    of chemical and thermal freeze-out points. The legend refers to
    the symbols for the thermal freeze-out points. For the original
    references for all the data points see \cite{CR99}.}

My first goal is a critical discussion of how these freeze-out 
parameters were extracted from the data and how reliable 
Fig.~1 is. Following that is a more detailed study of the 
fireball properties at chemical freeze-out, taking into account 
additional information not contained in Fig.~1, and a 
discussion of a consistent dynamical picture which can explain
Fig.~1. My main conclusion, based on a chain of arguments 
developed and sharpened over the last few years \cite{UH}, is
given in the abstract; similar conclusions were reached and
recently publicized by R. Stock \cite{Stock} and are also found in 
E. Shuryak's talk \cite{Shuryak}. 

\vspace*{-0.1cm}
\section{Thermal freeze-out, ``Hubble''-flow, and the Little Bang}
\vspace*{-0.05cm}

Let me begin with a discussion of the thermal freeze-out points. 
Freeze-out marks the transition from a strongly coupled system, which
evolves from one state of local thermal equilibrium to another, to
a weakly coupled one of essentially free-streaming particles. If this 
transition happens quickly enough, the thermal momentum distributions 
(superimposed by collective expansion flow) are frozen in, and the 
temperature and collective flow velocity at the transition ``point'' 
can be extracted from the measured momentum spectra. In high energy 
heavy-ion collisions the freeze-out process is triggered dynamically
by the accelerating transverse expansion and the very rapid growth
of the mean free paths as a result of the fast dilution of the matter 
\cite{SH94}. Idealizing the kinetic freeze-out process by a single 
point in the phase diagram is therefore not an entirely unreasonable 
procedure.

As in the Big Bang, the observed momentum spectra mix the thermal
information with the collective dynamics of the system. In the Big 
Bang, the observed microwave background radiation has a Bose-Einstein
energy spectrum with an ``effective temperature'' (inverse slope) 
which is {\em redshifted} by cosmological expansion down from the 
original freeze-out temperature of about 3000 K to an observed value 
of only 2.7 K. In the Little Bang, where we observe the {\em thermal 
hadron radiation} from the outside, the transverse momentum spectra 
are {\em blueshifted} by the collective transverse motion towards 
the observer. Simple approximate expressions which capture this effect 
are\footnote{This is accurate for non-relativistic particles from a 
Gaussian source with a linear transverse velocity profile $v_\perp(r)
 = \langle v_\perp \rangle {r\over r_{\rm rms}}$, where $\langle 
 v_\perp \rangle$ is the radial velocity at the rms radius 
$r_{\rm rms}^2{=}\langle x^2{+}y^2\rangle$. The analogous formula 
in \cite{CL96} lacks the factor ${1\over 2}$ in the second term 
since it uses the radial velocity at $r=r_{\rm rms}/\sqrt{2}$.
\label{FN1}}
$T_{\rm slope} \approx T_{\rm therm} + {1\over 2} m \langle v_\perp
 \rangle^2$ (which applies for $p_\perp{\ll}m$) and $T_{\rm slope}
 \approx T_{\rm therm} \sqrt{(1{+}\langle v_\perp \rangle)/
 (1{-}\langle v_\perp \rangle)}$ (which is good for $p_\perp{\gg}m$). 
For a given species (fixed $m$) the measured slope of the spectrum 
is thus ambiguous: temperature and flow cannot be separated. 

The ambiguity can be lifted in two ways: (i) One performs a simultaneous 
fit of the $m_\perp$-spectra of hadrons with different rest masses, 
thereby exploiting the mass dependence in the first of these two 
expressions \cite{Xu}. This makes the implicit assumption
that thermal freeze-out happens simultaneously for all particle
species. For Au+Au collisions at the AGS this works well and gives 
$T_{\rm therm}\approx 93$ MeV and $\langle v_\perp \rangle \approx 0.5\,c$
at midrapidity \cite{DSH99}. Or (ii) one concentrates on a single
particle species and correlates (as described in detail by
U.~Wiede\-mann \cite{Wqm99}) their spectra with their two-particle
Bose-Einstein correlations. The $M_\perp$-dependence of the transverse
HBT radius parameter\footnote{$\xi\sim{\cal O}(1)$ accounts for 
different transverse density and flow profiles; $\xi={1\over2}$ for 
the case described in fn.~\ref{FN1}. \label{FN2}} 
$R_\perp(M_\perp) \approx R/\sqrt{1+ \xi \langle v_\perp \rangle^2 
 {M_\perp\over T_{\rm therm}}}$ then provides an orthogonal correlation 
between temperature and flow, allowing for their separation. For pions 
from Pb+Pb collisions at the SPS this leads again to kinetic freeze-out 
temperatures of 90--100 MeV and average tranverse flow velocities of 
0.5--0.55\,$c$ (perhaps even somewhat higher at midrapidity) 
\cite{Wqm99,NA49}. 

This fixes the position of the freeze-out point along the $T$-axis in 
Fig.~1, but what about $\mu_{\rm B}$? Since chemical equilibrium 
is already broken earlier, at $T_{\rm chem} \approx 170-180$ MeV (see 
below), $\mu_{\rm B}$ is, strictly speaking, not well-defined. In order
to still be able to associate with kinetic freeze-out a point in the 
phase-diagram one commonly adjusts $\mu_{\rm B}$ in such a way that the
deviations between the observed particle ratios and their chemical
equilibrium values at $T_{\rm therm}$ are minimized. This is 
acceptable if the deviations are small; in practice they can 
approach a factor 2 or so. This clearly causes irreducible systematic 
uncertainties in the baryon chemical potential at thermal freeze-out 
which are usually not evaluated and are not included in the horizontal 
error bars in Fig.~1 (where given).

Let's nonetheless accept these thermal freeze-out parameters and now 
ask the question: {\em How did the system get there?} Does the implied 
picture of a rapidly expanding, locally thermalized fireball, 
{\em the Little Bang}, make sense? These questions can be studied 
by microscopic kinetic simulations of RQMD, URQMD and HSD \cite{Sqm} 
type; even if they do not include quark-gluon degrees of freedom 
during the very dense initial stages and thus may parametrize the 
initial hadron production incorrectly (see below), they can be used 
to explore the effects of scattering among the hadrons before 
kinetic freeze-out and the evolution of collective flow. A detailed 
study of thermalization by rescattering was recently performed by the 
URQMD group for Au+Au and Pb+Pb collisions from AGS to SPS energies 
\cite{Bravina} (Fig.~2). After an initial non-equilibrium stage 
lasting for about 8--10 fm/$c$ these systems reach a state of 
approximate local thermal equilibrium which expands and cools at 
roughly constant entropy for another 10 fm/$c$ before decoupling. 
During the adiabatic expansion stage strong collective flow builds 
up. Thermalization is driven by intense elastic rescattering, 
dominated by resonances (e.g. $\pi{+}N{\to}\Delta{\to}\pi{+}N$); 
inelastic processes are much rarer and lead only to minor changes 
in the chemical composition of the fireball \cite{Bass,Soff}. As a 
result, significant deviations from chemical equilibrium occur which 
increase with time; most importantly, at thermal freeze-out one 
sees a large pion excess which can only partially be accounted for 
by the initial string fragmentation process \cite{Bravina}. 
Remarkably, these deviations from chemical equilibrium produce very 
little entropy \cite{Bravina}. The $S/B$ values extracted from the 
URQMD simulations agree with those from the thermal model analysis 
of the data (cf. Figs.~1 and 2).

\begin{minipage}[c]{10cm}
\label{F2}
\vspace*{0.5cm}
\hspace*{-0.5cm}
  \epsfxsize 8cm
  \epsfbox{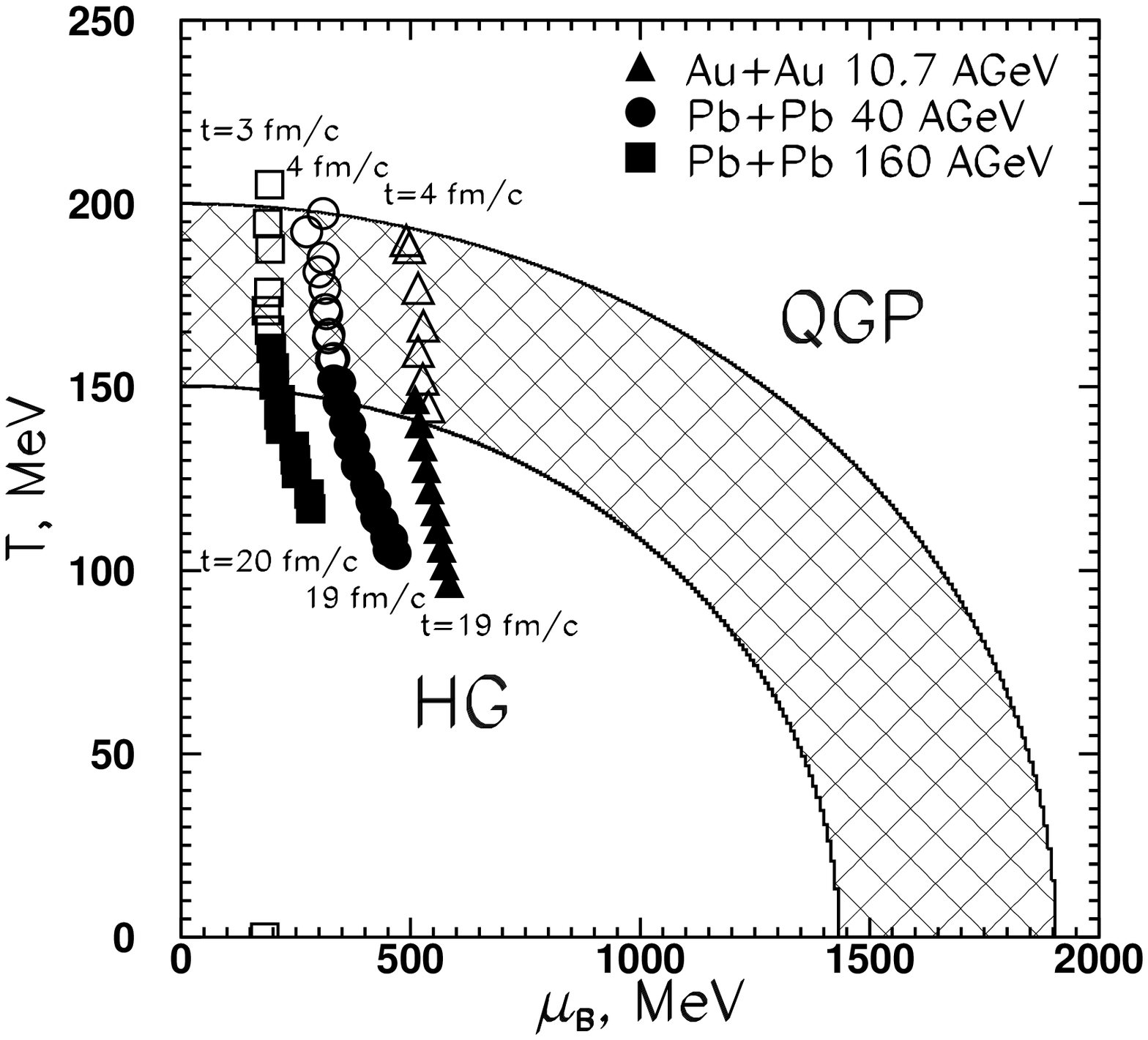}
\end{minipage}
\hfill\hspace*{-1.5cm}
\parbox[c]{6.5cm}{\vspace*{-0.3cm}
    Figure~2. Expansion trajectories from URQMD simulations
    \cite{Bravina}. Open and closed symbols denote the pre-equilibrium
    and hydrodynamic stages, respectively, of the collision in steps
    of 1 fm/$c$. The filled symbols lie on lines of constant entropy
    per baryon, $S/B$=38, 20, 12 for 160, 40, 10.7 $A$ GeV,
    respectively \cite{Bravina}. The shaded region indicates the
    expected parameter range for the deconfining phase transition.}
\vspace*{0.2cm}

Hadron momentum spectra and two-particle correlations thus provide 
strong evidence for the existence of the Little Bang: thermal hadron 
radiation with $T_{\rm therm}\approx 90-100$ MeV and strong 3-dimensional
(``Hubble-like'') expansion with transverse flow velocities 
$\langle v_\perp \rangle \approx 0.5-0.55\,c$ (and even larger 
longitudinal ones \cite{SH94,DSH99}). These two observations play a
similar role here as the discovery of Hubble expansion and the 
cosmic microwave radiation played for the Big Bang. But is there 
also a heavy-ion analogy to primordial Big Bang nucleosynthesis?
In the following section I will argue that we have indeed evidence 
for ``primordial hadrosynthesis'' in the Little Bang.

\vspace*{-0.1cm}
\section{Thermal models for chemical freeze-out and ``primordial
hadrosynthesis''}

Chemical reactions, which exploit small inelastic fractions of the 
total cross section, are typically much slower than the (resonance 
dominated) elastic processes. One thus expects chemical freeze-out 
to occur {\em before} thermal freeze-out \cite{H93} ($T_{\rm chem}
 > T_{\rm therm}$) but on an expansion trajectory with roughly the 
same entropy per baryon $S/B$. Fig.~1 suggests that this is indeed 
the case. This is analogous to the Big Bang where nucleosynthesis 
happened after about 3 minutes at $T_{\rm chem} \approx 100$ keV 
whereas the microwave background decoupled much later, after about 
300000 years at $T_{\rm therm} \approx {1\over 4}$ eV. The much 
smaller difference between the two decoupling temperatures in the 
Little Bang is mainly due to its much (about 18 orders of magnitude) 
faster expansion rate.

Before discussing implications of the chemical freeze-out points in 
Fig.~1 I first explain how they were obtained. Can thermal models 
be used to analyze chemical freeze-out? I discussed this question 
in some detail last year in Padova \cite{UH} and thus will be short 
here. The first difficulty arises from the collective expansion 
which strongly affects the shape of the $m_\perp$- and $y$-spectra, 
in a way which depends on the particle mass. However, many
experiments measure the particle yields only in small windows of
$m_\perp$ and $y$. A chemical analysis of particle ratios from
such experiments depends very strongly on model assumptions about
the fireball dynamics \cite{SSH95}. Static fireball fits yield 
chemical freeze-out parameters which are quite sensitive to the 
rapidity window covered by the data \cite{SHSX99}. Flow effects 
drop out, however, from $4\pi$-integrated particle ratios as long 
as freeze-out occurs at constant $T$ and $\mu$. $4\pi$ yields thus 
minimize the sensitivity to the collective fireball dynamics and are 
preferable for thermal model analyses.

\begin{minipage}[c]{10.5cm}
\label{F3}
\vspace*{-0.7cm}
\hspace*{-0.7cm}
  \epsfxsize 11.7cm
  \epsfbox{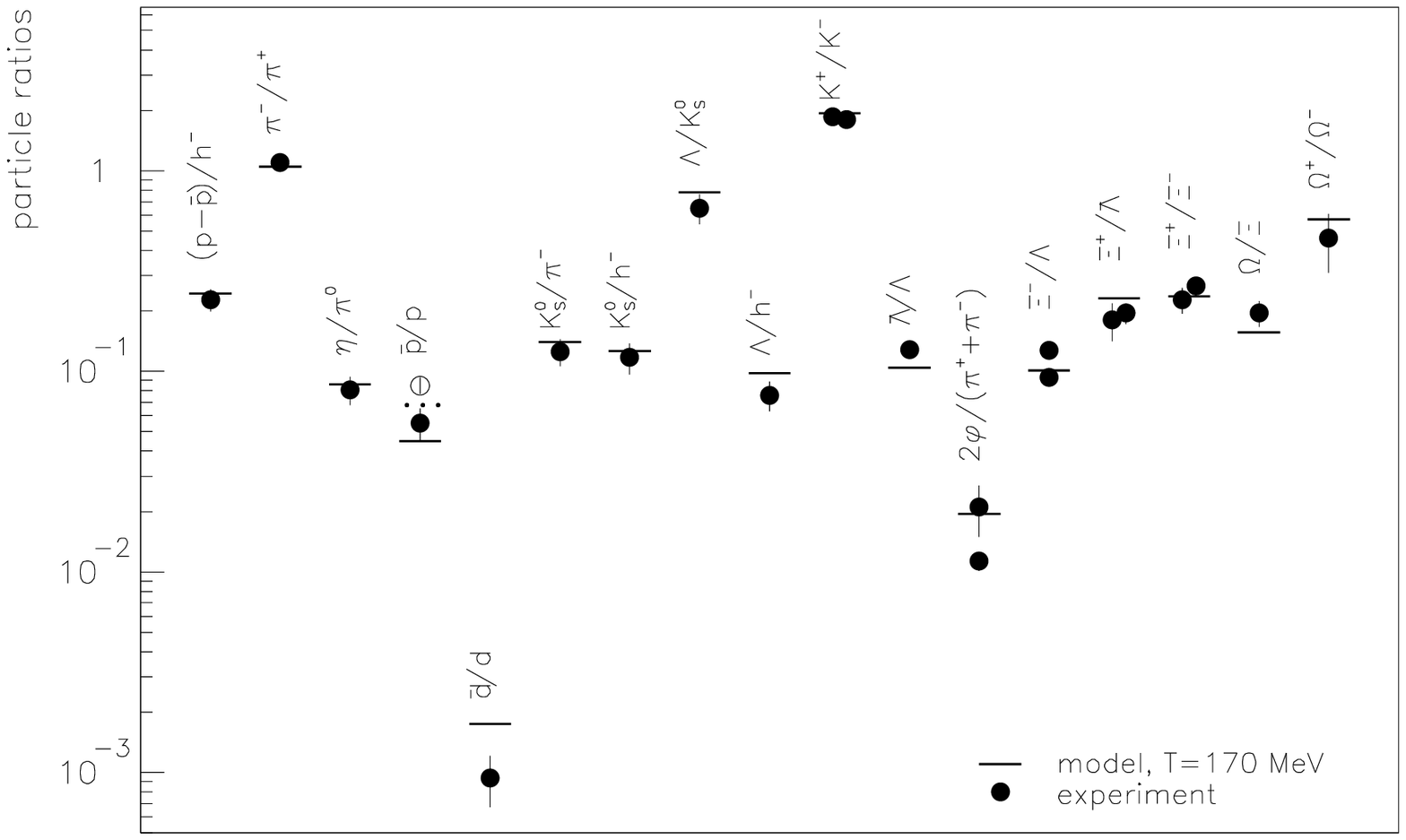}
\end{minipage}
\hfill\hspace*{-1.5cm}
\parbox[c]{5.1cm}{\vspace*{-0.7cm}
    Figure~3. Comparison between thermal model predictions
    and data for 158 $A$ GeV Pb+Pb collisions, after optimizing the
    model parameters $T_{\rm chem}$=170 MeV, $\mu_{\rm B}$=270 MeV,
    $\gamma_{\rm s}$=1 \cite{PBM}. Discrepancies between model and
    data remain below the systematic uncertainties of the model and
    among different data sets.}
\vskip -0.8cm

Of course, dynamic systems never freeze out at constant $T$ and 
$\mu$. While the steep $T$-dependence of the particle densities 
(which determine the local scattering rates and control freeze-out 
\cite{BGZ78}) prohibits strong temperature variations across the 
freeze-out surface \cite{MH97}, incomplete baryon number stopping
causes at higher energies significant longitudinal variations of 
$\mu_{\rm B}$ \cite{SSH95}. A global thermal fit replaces the 
freeze-out distributions $T_{\rm chem}(x)$, $\mu_{\rm B}(x)$ by 
average values $\langle T_{\rm chem} \rangle$, $\langle \mu_{\rm B}
 \rangle$. A recent study by Sollfrank \cite{S99}, in which he 
performed a global thermal fit to particle yields from a hydrodynamic 
calculation, showed that after optimizing $\langle T_{\rm chem}
 \rangle$ and $\langle \mu_{\rm B} \rangle$ the thermal model 
predicted yields which differed by up to 15\% from the actual 
ones, although hydrodynamic simulations implement perfect (local) 
chemical equilibrium by construction. {\em Thermal model fits can thus
never be expected to be perfect}; without detailed dynamical 
assumptions local variations of the thermal parameters in the real 
collision can never be fully absorbed by the model. Discrepancies 
between model and data at the 15-30\% level are inside the systematic
uncertainty band of the thermal model approach. While being impressed 
by how thermal models can reproduce the measured particle ratios at 
this level of accuracy over up to 3 orders of magnitude (see Fig.~3), 
I am thus deeply suspicious of ``perfect'' thermal model fits.

A second lesson to be learned from the exercise in \cite{S99} is that 
the $\chi^2$/d.o.f. resulting from such a fit is not very useful 
as an absolute measure for the quality of the fit: since 
discrepancies between the real yields and the predictions from 
the global thermal model cannot be avoided, $\chi^2$/d.o.f. becomes 
larger and larger as the data become more and more accurate. While 
$\chi^2$ minimization can still be used to identify the optimal model 
parameters within a given model, one should be very careful in 
using the absolute value of $\chi^2$/d.o.f. to judge the relative 
quality of different model fits \cite{RL99}. 

Let me note that the {\em ideal system for thermal model fits of 
particle ratios} will be provided by heavy-ion collisions at RHIC 
and LHC: hadron formation will happen at the confinement transition, 
and near midrapidity the baryon density is so low that $T_{\rm c}$ 
is nearly independent of $\mu_{\rm B}$. Replacing $T(x)$ by 
$\langle T\rangle$ will then be an excellent approximation. Due 
to longitudinal boost-in\-va\-ri\-ance near midrapi\-di\-ty, 
know\-ledge of $dN/dy$ will be good enough for a reliable chemical 
ana\-ly\-sis. Finally, transverse flow can only be stronger at RHIC 
and LHC than at the SPS, so freeze-out will happen even more quickly 
after hadronization, strengthening the primordial character of the 
observed particle ratios. 

\begin{minipage}[c]{10cm}
\label{F4}
\hspace*{-0.5cm}
  \epsfxsize 9cm
  \epsfbox{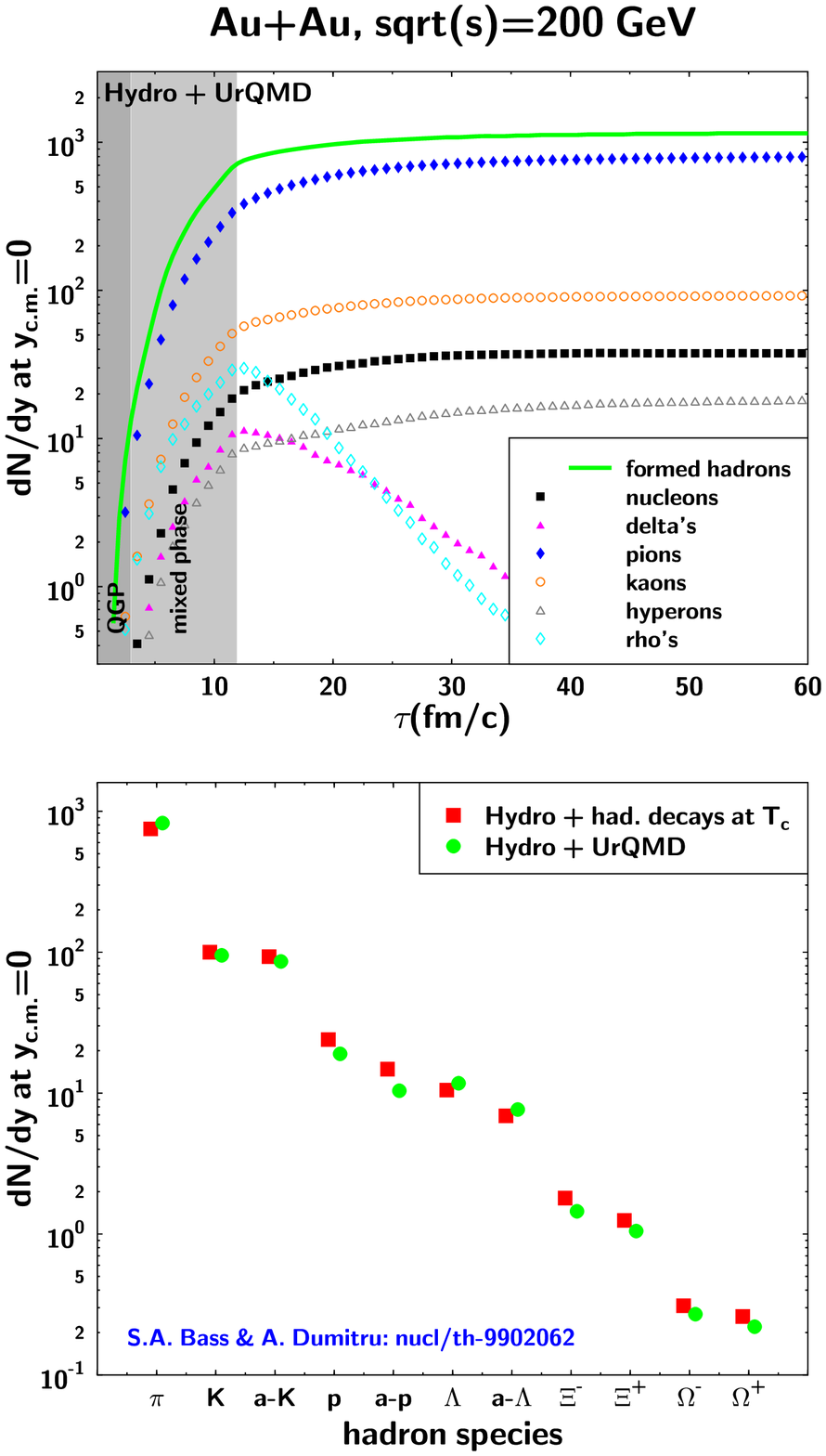}
\end{minipage}
\hfill\hspace*{-1cm}
\parbox[c]{6.3cm}{\vspace*{0.1cm}
    Figure~4.\\
    Upper part: time dependence of midrapidity hadron densities 
    for Au+Au colli\-sions at RHIC, calculated in a combined 
    hydrodynamic + URQMD simulation \cite{Bass}. At the 
    ha\-dro\-ni\-za\-tion temperature $T_{\rm c}$ hadrons are 
    created from the hydrodynamic phase with chemical equilibrium 
    abundances and are then evolved kinetically by URQMD. \\
    Lower part: final hadron abundances at the end of the kinetic 
    stage (circles) and if the calculation is stopped and all 
    re\-so\-nances are decayed directly at $T_{\rm c}$ (squares).}

These expectations are borne out by a recent analysis by Bass and
Dumitru \cite{Bass} who combined a hydrodynamic description of the
dense early stage with a URQMD simulation of the late hadronic stage. 
Fig.~4 (bottom) shows that indeed chemical freeze-out occurs quickly 
after hadronization: the yields at hadronization (squares) and after 
the last elastic scattering (circles) differ by less than 30\%, 
in spite of many collisions in between \cite{Bass}.

The fit of Pb+Pb data in Fig.~3 \cite{PBM} yields a chemical 
freeze-out temperature $T_{\rm chem} \approx 170$ MeV at full 
strangeness saturation ($\gamma_{\rm s}$=1). In
\cite{RL99} the same data are fit with $T_{\rm chem}=144\pm 2$ MeV 
and a strongly oversaturated strange phase-space ($\gamma_{\rm s}
 = 1.48\pm 0.08$). The authors also allow for oversaturation of the 
light quarks and find $\gamma_{\rm q}{=}1.72{\pm}0.08$ which allows
to absorb the large pion multiplicity at a low value of 
$T_{\rm chem}$. This ``chemical non-equilibrium'' \cite{RL99} fit 
underpredicts $\bar\Omega/\bar\Xi$ by 40\% and $\Omega/\Xi$ by 60\%,
a problem which disappears at $T_{\rm chem}{=}170$ MeV (Fig.~3). More 
importantly, since $\gamma_{\rm q}^2=e^{\mu_\pi/T}$, the freeze-out 
parameters in \cite{RL99} imply a very large pion chemical potential 
$\mu_\pi=156$\,MeV\,$>m_\pi$; this invalidates the assumption 
\cite{RL99} that Bose statistics for pions can be accounted for 
by considering only the first correction to the Boltzmann term. Hence, 
while the authors of \cite{RL99} prefer their fit on the basis of 
a low $\chi^2$/d.o.f., it has systematic uncertainties which far 
exceed the statistical errors given in \cite{RL99}. 

Having established the location in the phase diagram where chemical 
freeze-out occurs, we should again ask: {\em How did the system get 
there?} Since $T_{\rm chem}$ turns out to be very close to the 
predicted critical value for the hadronization phase transition, 
there is clearly no time between hadron formation and chemical 
freeze-out for kinetic equilibration of the hadron abundances by
inelastic hadronic rescattering \cite{Stock}. The observed hadronic 
chemical equilibrium at $T_{\rm chem}$ must therefore be 
{\em pre-established}: it reflects a statistical occupation of the 
hadronic phase-space, following the principle of maximum entropy, by
the hadronization process \cite{UH}. Hadrons form from a prehadronic
stage by filling each available phase-space cell with equal 
probability, subject only to the constraints of energy, baryon number
and strangeness conservation (the latter includes a possible overall
suppression of strangeness). Afterwards, the chemical composition
decouples essentially immediately, without major modifications by 
hadronic rescattering.

The parameter $T_{\rm chem}$ is thus {\em not a hadronic temperature 
in the usual sense}, i.e. not a result of hadronic kinetic 
equilibration. In the maximum entropy spirit it should be interpreted 
as a Lagrange multiplier which regulates the hadron abundances in 
accordance with the conservation laws and is directly related
to the critical energy density at which hadronization can proceed. 
$T_{\rm chem}{\approx}170$ MeV translates into $\epsilon_{\rm c}{\approx}1$ 
GeV/fm$^3$. This also explains naturally Becattini's observation 
\cite{Bec,UH} of hadronic chemical equilibrium at the same value of 
$T_{\rm chem}$ in $e^+e^-$, $pp$ and $p\bar p$ collisions at 
essentially all collision energies, although there hadronic final state 
interactions are completely absent. Whether the constituents of the 
prehadronic stage themselves thermalize before or during hadronization 
is a question which final hadron abundances cannot answer; as likely as 
their thermalization may appear, it is not necessary for an explanation 
of the observed phenomena.

The concept of statistical hadron formation from a pre-existing 
state of color-deconfined, completely uncorrelated quarks and 
antiquarks is supported by a recent analysis\footnote{For an 
improved argument taking into account global flavor conservation 
during hadronization see \cite{Zim}.}
of Bialas \cite{Bial} which follows similar earlier arguments by 
Rafelski \cite{Raf} but formulates them more generally such that
they do not require thermalization. Bialas points out that by 
consi\-de\-ring baryon/antibaryon (or generally particle/antiparticle)
ratios, the unknown effects on hadronization from the internal hadron 
structure drop out and one can check directly whether the observed 
hadron abundances can be fully understood by just counting their
conserved quantum numbers (carried by their valence quarks), or whether
additional correlations among the quarks exist. He finds \cite{Bial}
that in S+S and Pb+Pb collisions at the SPS the former is true while 
hadron production yields in p+Pb collisions point to correlations 
among the quarks.

\vspace*{-0.1cm}
\section{Early memories: strangeness enhancement}

The one decisive feature which distinguishes heavy-ion from elementary 
particle collisions is the strangeness content in the hadronic final 
state: the global strangeness fraction of the produced quark-antiquark 
pairs, $\lambda_{\rm s} = 2 \langle s\bar s \rangle/
 \langle u\bar u{+}d\bar d \rangle\vert_{\rm produced}$, is about 2 
times higher in nuclear collisions \cite{Odyn,BGS}. This can not be
reproduced by hadronic rescattering models \cite{Odyn,Soff} and must 
thus be a feature of the prehadronic state. Here I would like to 
discuss in more detail the specific enhancement factors for 
$K,\bar K, \Lambda,\bar\Lambda,\Xi,\bar\Xi, \Omega,\bar\Omega$, and 
$\phi$ reported recently and during this meeting \cite{senh,senh_qm99}.  

\begin{minipage}[c]{11cm}
\label{F5}
\vspace*{-0.3cm}
\hspace*{-0.5cm}
  \epsfxsize 11cm
  \epsfbox{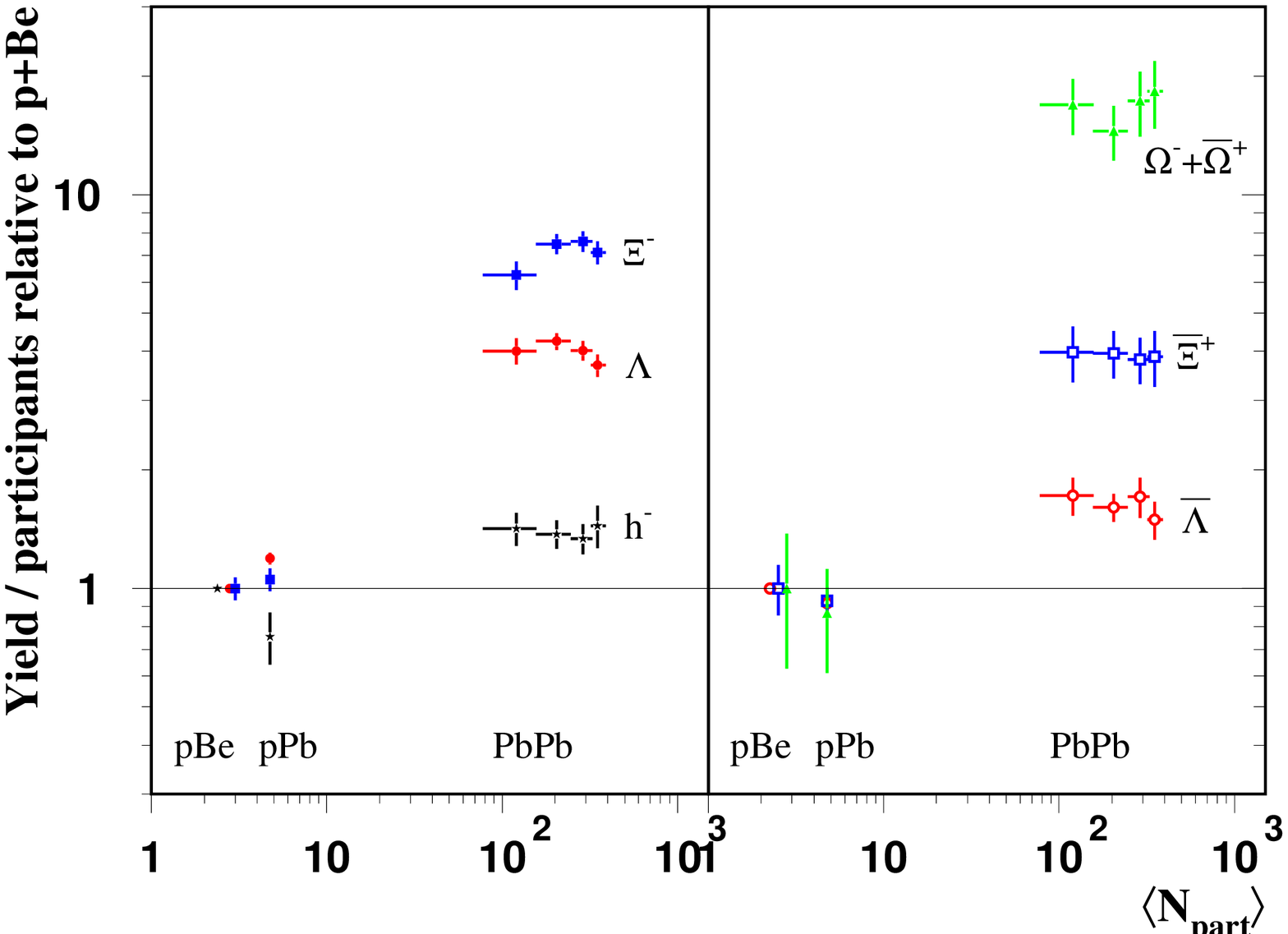}
\end{minipage}
\hfill\hspace*{-1.5cm}
\parbox[c]{5.3cm}{\vspace*{-0.5cm}
    Figure~5. Centrality dependence of strangeness enhancement as 
    mea\-su\-red by WA97 \cite{senh,senh_qm99}. 
    The strange parti\-cle yields per par\-ti\-ci\-pa\-ting 
    nucleon in Pb+Pb collisions at the SPS are compared 
    to the same ratio in p+Be and p+Pb collisions.}
\vspace*{0.2cm}

Fig.~5 shows that the relative strangeness enhancement between
Pb+Pb and p+Be collisions is the stronger the more strange quarks the
hadron contains. This is perfectly consistent with the above picture
of statistical hadronization: disregarding other phase-space 
constraints, an $\Omega$, for example, which contains 3 strange 
quarks, is expected to be enhanced by a factor $2^3{=}8$ if strange 
quarks are enhanced by a factor 2. On the other hand, this pattern
contradicts expectations from final state hadronic rescattering:
since hadrons with more strange quarks are heavier and strangeness
must be created in pairs, the production of stranger particles is 
suppressed by increasingly higher thresholds. 

An interesting observation from Fig.~5 is the apparent centrality 
independence of the specific strangeness enhancement factors: the 
enhancement appears to be already fully established in semiperipheral 
Pb+Pb collisions with about 100 participating nucleons. In fact, the 
global enhancement by a factor 2 was already seen in S+S collisions 
by NA35 \cite{Odyn}. Since it must be a prehadronic feature, but 
the lifetime of the prehadronic stage is shorter for smaller collision 
systems, this points to a new fast strangeness production mechanism 
in the prehadronic stage. Exactly this was predicted for the 
quark-gluon plasma \cite{RM82}.

At this meeting we saw new data on the centrality dependence of 
hadron yields \cite{senh_qm99}. Unfortunately, different 
centrality measures and prescriptions to determine $N_{\rm part}$
have been used. This needs clarification before the pattern in 
Fig.~5 can be considered confirmed.

Much recent effort went into trying to explain these observations
within microscopic simulations based on string breaking followed by 
hadronic rescattering. {\em All such attempts failed.} VENUS and RQMD 
give more strangeness enhancement for more central collisions 
\cite{Caliandro}, however not from hadronic rescattering, but 
mostly from the non-linear rise of the formation probability for 
quark matter droplets and color ropes. Strangeness enhancement is 
thus put in as an initial condition; unlike Fig.~5 it rises 
monotonically with $N_{\rm part}$. HIJING/$B\bar B$ \cite{Vance} 
uses baryon junction loops to enhance strange baryon production 
near midrapidity. Again this puts the enhancement into the initial 
conditions. The measured $N_{\rm part}$-dependence is not reproduced. 
The model also disagrees with the observed pattern $\bar\Omega/\Omega
 > \bar \Xi /\Xi > \bar \Lambda/\Lambda$ (which the statistical 
hadronization picture \cite{Bial} explains nicely). Finally, the 
``improved dual parton model'' \cite{Cap}, which gets some fraction 
of the enhancement in the initial state from ``diquark-breaking 
collisions'' and claims to obtain an even larger additional 
enhancement from hadronic final state interactions, suffers from 
a severe violation of detailed balance: it only includes inelastic 
channels (like $\pi{+}\Xi{\to}\Omega{+}K$) which increase multistrange 
baryons but neglects the (at least equally important \cite{KMR}) 
strangeness exchange processes (like $\pi{+}\Omega{\to}\bar K{+}\Xi$) 
which destroy them. Consequently it also fails to reproduce the 
apparent saturation of the enhancement factors seen in Fig.~5.

The observed strangeness enhancement pattern thus cannot be generated 
by hadronic final state interactions, but must be put in at the
beginning of the hadronic stage. No working model which does so in 
agreement with the data is known, except for conceptually simplest 
one, the statistical hadronization model.

\vspace*{-.15cm}
\section{Summary}
\vspace*{-0.05cm}

The analysis of soft hadron production data at the SPS indicates 
that hadron formation proceeds by statistical hadronization from a 
prehadronic state of uncorrelated (color-deconfined) quarks. This 
leads to pre-established apparent chemical equilibrium among the 
formed hadrons at the confinement tempe\-ra\-ture $T_{\rm c}$; it 
is not caused by kinetic equilibration through hadronic rescattering. 
After hadronization the hadron abundances freeze out more or less 
immediately. The chemical freeze-out temperature thus coincides 
with the critical temperature, $T_{\rm chem} \approx T_{\rm c}
 \approx 170$-180 MeV, corresponding to a critical energy density
$\epsilon_{\rm c}\approx 1$ GeV/fm$^3$ as predicted by lattice QCD.

The prehadronic state in $A+A$ ($A{\geq}32$) collisions contains about
twice more strangeness than in $e^+e^-$ and $pp$ collisions. This 
strangeness enhancement appears to be already fully established in
nuclear collisions with 60 or more participant nucleons and can not be 
ge\-ne\-rated by hadronic final state interactions. This suggests a fast 
$s\bar s$ creation mechanism in the prehadronic stage, as predicted 
for a quark-gluon plasma.

A clear hierarchy between chemical ($T_{\rm chem}\approx 170$--180 MeV) 
and thermal ($T_{\rm therm}\approx 90$--100 MeV) freeze-out is observed
in Pb+Pb collisions at the SPS; the gap is somewhat smaller 
($\approx 130$--140 MeV vs. $\approx 90$ MeV) at the AGS. In both cases 
thermal decoupling is accompanied by strong radial collective flow.
The smaller inverse slopes of the $\Omega$ $m_\perp$-spectra suggest
that a considerable fraction (but probably not all) of this flow
is generated by strong elastic hadronic rescattering after hadronization
\cite{vH}.

I conclude that we have seen the Little Bang in the laboratory, and that
most likely it is initiated by a quark-gluon plasma. 

{\em Acknowledgements:} I thank S.~Bass, J.~Cleymans and 
R.~Lietava for providing me with Figs.~1, 4, and 5. Several fruitful 
discussions with R.~Stock are gratefully acknowledged. A comment by 
Zhangbu Xu led to the clarification presented in Footnote 1.
\vspace*{-0.3cm}



\begin{thebibliography}{99}
\bibitem{CR99} 
  J. Cleymans, K. Redlich, Phys. Rev. Lett. 81 (1998) 5284; and
  nucl-th/9903063.
\bibitem{UH}
  F. Becattini, U. Heinz, Z. Phys. C76, (1997) 269; 
  U. Heinz, Nucl. Phys. A638 (1998) 357c; 
            J. Phys. G25 (1999) 263;
            and hep-ph/9902424.
\bibitem{Stock}
  R. Stock, Prog. Part. Nucl. Phys. 42 (1999) 295; 
  Phys. Lett. B456 (1999) 277.
\bibitem{Shuryak}
  E. Shuryak, this volume (hep-ph/9906443).
\bibitem{SH94}
  E. Schnedermann, U. Heinz, Phys. Rev. C50 (1994) 1675.
\bibitem{CL96}
  T. Cs\"org\H{o}, B. L\"orstad, Phys. Rev. C54 (1996) 1390; 
  R. Scheibl, U. Heinz, Phys. Rev. C59 (1999) 1585.  
\bibitem{Xu}
  N. Xu et al. (NA44 Coll.), Nucl. Phys. A610 (1999) 175c.
\bibitem{DSH99} 
  H. Dobler, J. Sollfrank, U. Heinz, Phys. Lett. B457 (1999) 353.
\bibitem{Wqm99}
  U.A. Wiedemann, this volume, and references therein
  (nucl-th/9907048).
\bibitem{NA49}
  H. Appelsh\"auser et al. (NA49 Coll.), Eur. Phys. J. C2 (1998) 661.
\bibitem{Sqm}
  See contributions by H. Sorge, M. Bleicher, and W. Cassing in this volume.
\bibitem{Bravina}
  L. Bravina et al., J. Phys. G25 (1999) 351; hep-ph/9906548; 
  Phys. Lett. B, in press.
\bibitem{Bass}
  S. Bass, A. Dumitru et al., nucl-th/9902062.
\bibitem{Soff}
  S. Soff et al., nucl-th/9907026.
\bibitem{H93}
  U. Heinz, Nucl. Phys. A566 (1994) 563c; NATO ASI Series B346 (1995) 
  413. 
\bibitem{SSH95}
  C. Slotta, J. Sollfrank, U. Heinz, AIP Conf. Proc. 340 (1995) 462.
\bibitem{SHSX99}
  J. Sollfrank, U. Heinz, H. Sorge, N. Xu, Phys. Rev. C59 (1999) 1637.
\bibitem{BGZ78}
  J. Bondorf, S.I.A. Garpman, J. Zim\'anyi, Nucl. Phys. A296 (1978) 320.
\bibitem{MH97}
  E. Schnedermann, J. Sollfrank, U. Heinz, NATO ASI Series B303 (1993) 
  175; U.~Mayer, U. Heinz, Phys. Rev. C56 (1997) 439.
\bibitem{S99}
  J. Sollfrank, Eur. Phys. J. C9 (1999) 159.
\bibitem{PBM}
  P. Braun-Munzinger, I. Heppe, J. Stachel, nucl-th/9903010.
\bibitem{RL99}
  J. Rafelski, J. Letessier, nucl-th/9903018.
\bibitem{Bec}
  F. Becattini, Z. Phys. C69 (1996) 485.
\bibitem{Zim}
  J. Zim\'anyi, T. Bir\'o, P. L\'evai, hep-ph/9904501.
\bibitem{Bial}
  A. Bialas, Phys. Lett. B442 (1998) 449.
\bibitem{Raf}
  J. Rafelski, Phys. Lett. B262 (1991) 333.
\bibitem{Odyn}
  See discussion by G. Odyniec, Nucl. Phys. A638 (1998) 135c, and 
  references therein.
\bibitem{BGS}
  F. Becattini, M. Ga\'zdzicki, J. Sollfrank, Eur. Phys. J. C5 (1998) 143.
\bibitem{senh}
  E. Andersen et al. (WA97 Coll.), Phys. Lett. B433 (1998) 209;
  R. Lietava et al. (WA97 Coll.), J. Phys. G25 (1999) 181 (for the 
  correct Fig. 7 see p. 460 in the same volume);
  H. Appelsh\"auser et al. (NA49 Coll.), Phys. Lett. B444 (1998) 523.
\bibitem{senh_qm99}
  See F. Antinori, D. Elia (WA97 Coll.); F. Sikler, C. H\"ohne 
  (NA49 Coll.); and N. Willis (NA50 Coll.), this volume.
\bibitem{RM82}
  J. Rafelski, B. M\"uller, Phys. Rev. Lett. 48 (1982) 1066.
\bibitem{Caliandro}
  F. Antinori et al., Eur. Phys. J. C, in press; R. Caliandro,
  private communication.
\bibitem{Vance}
  S.E. Vance, M. Gyulassy, nucl-th/9901009.
\bibitem{Cap}
  A. Capella, C.A. Salgado, hep-ph/9903414.
\bibitem{KMR}
  P. Koch, B. M\"uller, J. Rafelski, Phys. Rep. 142 (1986) 167, Fig. 5.2.
\bibitem{vH}
  H. van Hecke, H. Sorge, N. Xu, Phys. Rev. Lett. 81 (1998) 5764.

\end{thebibliography}
\end{document}